\documentclass[12pt,journal, onecolumn, draftcls]{IEEEtran}
\usepackage{amsmath}
\usepackage{epsfig}
\usepackage{amssymb}
\begin{document}
\title{Transmit Energy Focusing for DOA Estimation in MIMO Radar
with Colocated Antennas}
%\ninept
\author{Aboulnasr~Hassanien,~\IEEEmembership{Member,~IEEE} and
Sergiy A.~Vorobyov~\IEEEmembership{Senior Member,~IEEE}

\thanks{This work is supported in parts by the Natural Science and
Engineering Research Council (NSERC) of Canada and the Alberta
Ingenuity Foundation, Alberta, Canada.

The authors are with the Department of Electrical and Computer
Engineering, University of Alberta, 9107-116 St., Edmonton,
Alberta, T6G~2V4 Canada. Emails: {\tt \{hassanie, vorobyov
\}@ece.ualberta.ca}

{\bf Corresponding author:} Sergiy A.~Vorobyov, Dept. Elect. and
Comp. Eng., University of Alberta, 9107-116 St., Edmonton,
Alberta, T6G 2V4, Canada; Phone: +1 780 492 9702, Fax: +1 780 492
1811. Email: {\tt vorobyov@ece.ualberta.ca}. }

}
\vspace{-1.0cm}
\maketitle
\vspace{-1.0cm}
%\begin{center}
%\centerline{ EDICS: SAM-SAMC}
%\end{center}
%\vspace{-0.5cm}

\begin{abstract}
In this paper, we propose a transmit beamspace energy focusing
technique for multiple-input multiple-output (MIMO) radar with
application to direction finding for multiple targets. The general
angular directions of the targets are assumed to be located within
a certain spatial sector. We focus the energy of multiple (two or
more) transmitted orthogonal waveforms within that spatial sector
using transmit beamformers which are designed to improve the
signal-to-noise ratio (SNR) gain at each receive antenna. The
subspace decomposition-based techniques such as MUSIC can then be
used for direction finding for multiple targets. Moreover, the
transmit beamformers can be designed so that matched-filtering the
received data to the waveforms yields multiple (two or more) data
sets with rotational invariance property that allows applying
search-free direction finding techniques such as ESPRIT for two
data sets or parallel factor analysis (PARAFAC) for more than two
data sets. Unlike previously reported MIMO radar
ESPRIT/PARAFAC-based direction finding techniques, our method
achieves the rotational invariance property in a different manner
combined also with the transmit energy focusing. As a result, it
achieves better estimation performance at lower computational
cost. Particularly, the proposed technique leads to lower
Cramer-Rao bound than the existing techniques due to the transmit
energy focusing capability. Simulation results also show the
superiority of the proposed technique over the existing
techniques.
\end{abstract}
%
%\vspace{-0.5cm}

\section{Introduction}
\label{sec:intro} The development of multiple-input
multiple-output (MIMO) radar is recently the focus of intensive
research \cite{MIMO_radar_book}-\cite{Radar_colocated_review}. A
MIMO radar is generally defined as a radar system with multiple
transmit linearly independent waveforms and it enables joint
processing of data received by multiple receive antennas. MIMO
radar can be either equipped with widely separated antennas
\cite{Radar_separated_review} or colocated antennas
\cite{Radar_colocated_review}.

Estimating direction-of-arrivals (DOAs) of multiple targets from
measurements corrupted by noise at the receiving array of antennas
is one of the most important radar applications frequently
encountered in practice. Many DOA estimation methods have been
developed for traditional single-input multiple-output (SIMO)
radar \cite{Pisarenko}-\cite{Haardt}. Among these methods the
estimation of signal parameters via rotational invariance
techniques (ESPRIT) and multiple signal classification (MUSIC) are
the most popular due to their simplicity and high-resolution
capabilities \cite{Schmidt}, \cite{Roy}, \cite{Haardt}. Moreover,
ESPRIT is a special and computationally efficient case of a more
general decomposition technique of high-dimensional (higher than
2) data arrays known as parallel factor analysis (PARAFAC)
\cite{Sidiropoulos1}, \cite{Vorobyov}.

More recently, some algorithms have been developed for DOA
estimation of multiple targets in the context of MIMO radar
systems equipped with $M$ colocated transmit antennas and $N$
receive antennas
\cite{Xu_Li_Stoica_2008}--\cite{PARAFAC_MIMO_radar_2009}. The
algorithms proposed in \cite{Xu_Li_Stoica_2008} and
\cite{Bekkerman} require an exhaustive search over the unknown
parameters and, therefore, mandate prohibitive computational cost
if the search is performed over a fine grid. On the other hand,
the search-free ESPRIT-based algorithms of
\cite{ESPRIT_MIMO_radar_2008} and \cite{ESPRIT_MIMO_radar_2008_1}
as well as PARAFAC-based algorithm of
\cite{PARAFAC_MIMO_radar_2009} utilize the rotational invariance
property of the so-called extended virtual array to estimate the
DOAs at a moderate computational cost. It is worth noting that in
the case of MIMO radar, the advantages of the aforementioned DOA
estimation methods over similar MUSIC- and ESPRIT/PARAFAC-based
DOA estimation methods for SIMO radar appear due to the fact that
the extended virtual array of $MN$ virtual antennas can be
obtained in the MIMO radar case by matched-filtering the data
received by the $N$-antenna receive array to $M$ transmitted
waveforms. Therefore, the effective aperture of the virtual array
can be significantly extended that leads to improved angular
resolution. In the methods of
\cite{ESPRIT_MIMO_radar_2008}--\cite{PARAFAC_MIMO_radar_2009}, the
rotational invariance property is achieved by partitioning the
receiving array to two (in ESPRIT case) or multiple (in PARAFAC
case) overlapped subarrays. Then, the rotational invariance is
also presumed for the virtual enlarged array of $MN$ virtual
antennas. However, the methods of
\cite{ESPRIT_MIMO_radar_2008}--\cite{PARAFAC_MIMO_radar_2009}
employ full waveform diversity, i.e., the number of transmitted
waveforms equals the number of transmit antennas, at the price of
reduced transmit energy per waveform. In other words, for fixed
total transmit energy, denoted hereafter as $E$, each waveform has
energy $E/M$ transmitted omni-directionally. It results in a
reduced signal-to-noise ratio (SNR) per virtual antenna. On the
other hand, it is well-known that the estimation accuracy of
subspace-based techniques suffers from high SNR threshold, that
is, the root-mean-squared error (RMSE) of DOA estimates approaches
the corresponding Cramer-Rao bound (CRB) only for relatively high
SNRs \cite{SN}, \cite{VanTrees}. Therefore, the higher the  SNR,
the better the DOA estimation performance can be achieved.

It has been shown in \cite{NasrSergConf}, \cite{NasrSerg} that a
tradeoff between the SNR gain and aperture of the MIMO radar
virtual array can be achieved by transmitting less than $M$
orthogonal waveforms. Exploiting this tradeoff, we develop in this
paper\footnote{An early exposition of a part of this work has been
presented in \cite{NasrSergCAMSAP}.} a group of DOA estimation
methods, which allow to increase the SNR per virtual antenna by
(i)~transmitting less waveforms of higher energy and/or
(ii)~focusing transmitted energy within spatial sectors where the
targets are likely to be located. At the same time, reducing the
number of transmitted waveform reduces the aperture of the virtual
array, while a larger aperture may be useful for increasing the
angular resolution at high SNR region.

The contributions of this work are based on the observation that
by using less waveforms the energy available for each transmitted
waveform can be increased, that is, the SNR per each virtual
antenna can be improved, while the aperture of the virtual array
decreases to $KN$, where $K\leq M$ is the number of orthogonal
waveforms. Moreover, the SNR per virtual antenna can be further
increased by focusing the transmitted energy in a certain sector
where the targets are located. Our particular contributions are as
follows. The transmit beamformers are designed so that the
transmitted energy can be focussed in a certain special sector
where the targets are likely to be located that helps to improve
the SNR gain at each receive antenna, and therefore, improve the
angular resolution of DOA estimation techniques, such as, for
example, MUSIC-based techniques. Moreover, we consider the
possibility of obtaining the rotational invariance property while
transmitting $1 < K \leq M$ orthogonal waveforms using different
transmit beamforming weight vectors and focusing the transmitted
energy on a certain spatial sector where the targets are located.
It enables us to design search-free ESPRIT/PARAFAC-based DOA
estimation techniques. In addition, we derive CRB for the
considered DOA estimation schemes that aims at further
demonstrating how the DOA estimation performance depends on the
number of transmitted waveforms, transmit energy focusing, and
effective array aperture of the MIMO radar virtual array.

The paper is organized as follows. In Section~II, MIMO radar
signal model is briefly introduced. In Section~III, we present the
transmit beamspace-based MIMO radar signal model. Two approaches
for designing the transmit beamspace weight matrix are given in
Section~IV. In Section~V, we present MUSIC and ESPRIT DOA
estimation for transmit beamspace-based MIMO radar. Performance
analysis and CRB are given in Section~VI. Simulation results which
show the advantages of the proposed transmit beamspace-based MIMO
radar DOA estimation techniques are reported in Section~VII
followed by conclusions drawn in Section~VIII.

\section{MIMO Radar Signal Model} \label{sec:MIMO_model}
Consider a MIMO radar system equipped with a transmit array of $M$
colocated antennas and a receive array of $N$ colocated antennas.
Both the transmit and receive antennas are assumed to be
omni-directional. The $M$ transmit antennas are used to transmit
$M$ orthogonal waveforms. The complex envelope of the signal
transmitted by the $m$th transmit antenna is modeled as
\begin{equation}\label{eq:emitted_signal}
s_m(t) = \sqrt{\frac{E}{M}}\phi_m(t),\quad m=1,\ldots,M
\end{equation}
where $t$ is the fast time index, i.e., the time index within one
radar pulse, $E$ is the total transmitted energy within one radar
pulse, and $\phi_m(t)$ is the $m$th baseband waveform. Assume that
the waveforms emitted by different transmit antennas are
orthogonal. Also, the waveforms are normalized to have
unit-energy, i.e., $\int_{T}|\phi_m(t)|^2dt=1,\, m=1,\ldots,M$,
where $T$ is the pulse width.

Assuming that $L$ targets are present, the $N \times 1$ received
complex vector of the receive array observations can be written as
\begin{eqnarray}\label{eq:target_signal}
{\bf x}(t, \tau) &\!\!\!=\!\!\!& \sum_{l=1}^L r_l(t,\tau){\bf b}
(\theta_l) + {\bf z}(t,\tau)
\end{eqnarray}
where $\tau$ is the slow time index, i.e., the pulse number, ${\bf
b}(\theta)$ is the steering vector of the receive array, ${\bf
z}(t,\tau)$ is $N\times 1$ zero-mean white Gaussian noise term,
and
\begin{eqnarray}\label{eq:reflected signal}
r_l(t,\tau)&\!\!\!\triangleq\!\!\!& \sqrt{\frac{E}{M}}
\alpha_l(\tau){\bf a}^{T}(\theta_l){\boldsymbol\phi}(t)
\end{eqnarray}
is the radar return due to the $l$th target. In (\ref{eq:reflected
signal}), $\alpha_l(\tau)$, $\theta_l$, and ${\bf a}(\theta_l)$
are the reflection coefficient with variance $\sigma^2_{\alpha}$,
spatial angle, and  steering vector of the transmit array
associated with the $l$th target, respectively,
${\boldsymbol\phi}(t)\triangleq[\phi_1(t),\ldots,\phi_M(t)]^T$ is
the waveform vector, and $(\cdot)^T$  stands for the transpose.
Note that the reflection coefficient $\alpha_l(\tau)$ for each
target is assumed to be constant during the whole pulse, but
varies independently from pulse to pulse, i.e., it obeys the
Swerling II target model \cite{PARAFAC_MIMO_radar_2009}.

Exploiting the orthogonality property of the transmitted
waveforms, the $N\times 1$ component of the received data
\eqref{eq:target_signal} due to the $m$th waveform can be
extracted using matched-filtering which is given as follows
\begin{equation}\label{eq:mthComponent}
{\bf x}_{m}(\tau)\triangleq \int_{T}{\bf
x}(t,\tau){\phi}_m^{*}(t)dt,\quad m=1,\ldots, M.
\end{equation}
where $(\cdot)^{*}$ is the conjugation operator. Stacking the
individual vector components \eqref{eq:mthComponent} in one column
vector, we obtain the $MN\times 1$ virtual data vector
\cite{Radar_colocated_review}
\begin{eqnarray}\label{eq:virtual_data}
{\bf y}_{\rm MIMO}(\tau) &\!\!\!\triangleq\!\!\!& [{\bf x}_1^T
(\tau) \cdots {\bf x}_{M}^T(\tau)]^T\nonumber\\
&\!\!\!=\!\!\!& \sqrt{\frac{E}{M}}\sum_{l=1}^{L}\alpha_l(\tau){\bf
a}(\theta_l)\otimes{\bf b}(\theta_l) + {\tilde{\bf z}}(\tau)\nonumber\\
&\!\!\!=\!\!\!& \sqrt{\frac{E}{M}}\sum_{l=1}^{L}\alpha_l(\tau)
{\bf u}_{\rm MIMO}(\theta_l) + {\tilde{\bf z}}(\tau)
\end{eqnarray}
where $\otimes$ denotes the Kronker product,
\begin{equation}\label{eq:MIMOvirtualSV}
{\bf u}_{\rm MIMO}(\theta)\triangleq{\bf a}(\theta)\otimes{\bf b}(\theta)
\end{equation}
is the $M N \times 1$ steering vector of the virtual array, and
${\tilde{\bf z}(\tau)}$ is the $MN\times 1$ noise term whose
covariance is given by $\sigma_z^2{\bf I}_{MN}$. The $M N \times
MN$ covariance matrix ${\bf R}_{\rm MIMO}\triangleq{\rm E}\{{\bf
y}_{\rm MIMO}(\tau){\bf y}^H_{\rm MIMO}(\tau)\}$ is hard to obtain
in practice. Therefore, the following sample covariance matrix
\begin{equation}\label{SampleCovMatrTxMIMO}
\hat{\bf R}_{\rm MIMO} = \frac{1}{Q} \sum_{\tau=1}^Q {\bf y}_{\rm
MIMO}(\tau){\bf y}^H_{\rm MIMO}(\tau)
\end{equation}
is used, where $Q$ is the number of snapshots.

\section{Transmit Beamspace Based MIMO Radar Signal Model}
\label{sec:TransmitBeamspace}

Instead of transmitting omin-directionally, we propose to focus
the transmitted energy within a sector $\boldsymbol\Theta$ by
forming $K$~directional beams where an independent waveform is
transmitted over each beam. Note that the spatial sector
$\boldsymbol\Theta$ can be estimated in a preprocessing step using
any low-resolution DOA estimation technique of low complexity.

Let ${\bf C}\!\triangleq[{\bf c}_1,\ldots,{\bf c}_K]^T$ be the
transmit beamspace matrix of dimension $M\times K\ (K\leq M)$,
where ${\bf c}_k$ is the $M\times1$ unit-norm weight vector used
to form the $k$th beam. The beamspace matrix can be properly
designed to maintain constant beampattern within the sector of
interest~$\boldsymbol\Theta$ and to minimize the energy
transmitted in the out-of-sector areas. Let ${\boldsymbol\phi}_K
(t) \triangleq [\phi_1(t), \ldots, \phi_K(t)]^T$ be the $K\times
1$ waveform vector. The $k$th column of ${\bf C}$ is used to form
a transmit beam for radiating the $k$th waveform $\phi_k(t)$. The
signal radiated towards a hypothetical target located at a
direction $\theta$ via the $k$th beam can be modeled as
\begin{equation}\label{eq:Tx_signal}
s_k(t, \theta) = \sqrt{\frac{E}{K}} \left( {\bf c}^H_k{\bf
a}(\theta) \right) {\phi}_k(t)
\end{equation}
where $\sqrt{E/K}$ is a normalization factor used to satisfy the
constraint that the total transmit energy is fixed to $E$. The
signal radiated via all beams towards the direction $\theta$ can
be modeled as
\begin{eqnarray}\label{eq:Tx_signals}
s(t, \theta) &\!\!\!\!\!=\!\!\!\!\!& \sqrt{\frac{E}{K}}
\sum^K_{k=1}  \left( {\bf c}^H_k{\bf
a}(\theta) \right) {\phi}_k(t) \nonumber\\
&\!\!\!\!\!=\!\!\!\!\!& \sqrt{\frac{E}{K}}{\bf C}^H{\bf a}
(\theta){\boldsymbol\phi}_K (t)= \sqrt{\frac{E}{K}}{\tilde{\bf
a}}^T (\theta){\boldsymbol\phi}_K (t)
\end{eqnarray}
where $(\cdot)^H$ stands for the Hermitian transpose and
\begin{equation}\label{eq:TxBeamManifold}
{\tilde{\bf a}}(\theta)\triangleq\left({\bf C}^H{\bf a}
(\theta)\right)^T.
\end{equation}
Then, the transmit beamspace can be viewed as a transformation
that results in changing the $M\times 1$ transmit array manifold
${\bf a}(\theta)$ into the $K\times 1$ manifold
\eqref{eq:TxBeamManifold}.

%It is worth noting that
%the beamspace energy, i.e., the energy of (\ref{eq:Tx_signals}) transmitted within
%one radar pulse is given by
%\begin{eqnarray}\label{eq:phased-MIMO_Ek}
%E_{\rm beam}\!&\!\!\!\!\!\triangleq\!\!\!\!\!&\!\int_{T}\!\|{\bf s}(t)\|^2dt\!=\!
%\frac{E}{K}\sum_{k=1}^K\|{\bf c}_k\|^2\!\!\int_{T}\!|{\phi}_k(t)|^2dt\!=\!E.
%\end{eqnarray}

At the receive array, the $N \times 1$ complex vector of array
observations can be expressed as
\begin{eqnarray} \label{eq:target_signal2}
{\bf x}_{\rm beam}(t, \tau)&\!\!\!\!\!=\!\!\!\!\!&
\sqrt{\frac{E}{K}}\!\sum_{l=1}^{L}\!\alpha_l(\tau)\left({\tilde{\bf a}}^T
(\theta_l){\boldsymbol\phi}_K(t)\right){\bf b}(\theta_l)\!+\!{\bf z}(t,\tau).
\end{eqnarray}
By matched-filtering ${\bf x}_{\rm beam}(t,\tau)$ to each of the
waveforms ${\phi_k}\ (k=1,\ldots, K)$, the received signal
component associated with each of the transmitted waveforms can be
obtained as
\begin{eqnarray}\label{eq:kthBeamComponent}
{\bf y}_{k}(\tau)&\!\!\!\triangleq\!\!\!& \int_{T}{\bf x}_{\rm beam}
(t,\tau){\phi}_k^{*}(t)dt\nonumber\\
&\!\!\!=\!\!\!&\sqrt{\frac{E}{K}}\!\sum_{l=1}^{L}\!\alpha_l(\tau)
{\tilde{\bf a}}_{[k]}(\theta_l){\bf b}(\theta_l)+ {\bf z}_k(\tau)
\end{eqnarray}
where $(\cdot)_{[k]}$ is the $k$th entry of a vector and the $N
\times 1$ noise term is defined as
\begin{equation}\label{eq:kthNoiseTerm}
{\bf z}_k(\tau) = \int_{T}{\bf z}(t,\tau){\phi}_k^{*}(t)dt.
\end{equation}

Stacking the individual vector components
\eqref{eq:kthBeamComponent} in one column vector, we obtain the
following $KN\times 1$ virtual data vector
\begin{eqnarray}\label{eq:virtual_data1}
{\bf y}_{\rm beam}(\tau) &\!\!\!\triangleq\!\!\!& [{\bf
y}_1^T(\tau) \cdots {\bf y}_{K}^T(\tau)]^T\nonumber\\
&\!\!\!=\!\!\!& \sqrt{\frac{E}{K}}\sum_{l=1}^{L}\alpha_l(\tau)
\left({\tilde{\bf a}}(\theta_l)\otimes{\bf b}(\theta_l)\right)
+ {\tilde{\bf z}}_K(\tau)
\end{eqnarray}
where ${\tilde{\bf z}_K(\tau)}\triangleq[{\bf z}_1(\tau), \ldots,
{\bf z}_K(\tau)]^T$ is the $KN\times 1$ noise term whose
covariance is given by $\sigma_z^2{\bf I}_{KN}$.

The transmit beamspace signal model given by
\eqref{eq:virtual_data1} provides the basis for optimizing a
general-shape transmit beampattern over the transmit beamspace
weight matrix ${\bf C}$. By carefully designing ${\bf C}$, the
transmitted energy can be focussed in a certain spatial sector, or
divided between several disjoint sectors in space. As compared to
traditional MIMO radar, the benefit of using transmit energy
focusing is the possibility to increase in the signal power at
each virtual array element. This increase in signal power is
attributed to two
factors:\\
(i) transmit beamforming gain, i.e., the signal power associated
with the $k$th waveform reflected from a target at direction
$\theta$ is magnified by factor $|{\bf c}^H_k{\bf a}(\theta)|^2$;\\
(ii) the signal power associated with the $k$th waveform is
magnified by factor $E/K$ due to dividing the fixed total transmit
power $E$ over $K\leq M$ waveforms instead of $M$ waveforms.

As it is shown in subsequent sections, the aforementioned two
factors result in increasing the SNR per virtual element which in
turn results in lowering the CRB and improving the DOA estimation
accuracy.

\section{Transmit Beamspace Design}
In this section, two methods for designing the transmit beamspace
weight matrix ${\bf C}$ are proposed.

\subsection{Motivations}
Given the beamspace angular sector-of-interest
$\boldsymbol\Theta$, several beamspace dimension reduction
techniques applied to the data at the output of a passive receive
array are reported in the literature (see \cite{HassGershWong} and
references therein). The essence of these beamspace dimension
reduction techniques is to perform DOA estimation in the reduced
dimension space rather than in the elementspace (full dimension of
received data) which leads to great computational savings.
Performing DOA estimation in a reduced dimension beamspace has
also proved to improve probability of source resolution as well as
estimation sensitivity \cite{Lee}. However, it is well known that
the CRB on DOA estimation performed in reduced dimension space is
higher than (or at best equal to) the CRB on DOA estimation
performed in elementspace, i.e., applying beamspace dimension
reduction techniques to passive arrays does not improve the best
achievable estimation performance \cite{Anderson}. This is natural
because beamspace dimension reduction in passive arrays just
preserves the signals observed within a certain spatial sector and
filters out signals observed outside that sector.

The transmit beamspace processing proposed in
Section~\ref{sec:TransmitBeamspace} focuses all transmit energy
within the desired spatial sector instead of spreading it
omni-directionally in the whole spatial domain. In other words,
the amount of energy that otherwise could be wasted in the
out-of-sector areas is added to the amount of energy to be
transmitted within the desired sector. As a result, the signal
strength within the desired sector is increased potentially
leading to improved best achievable estimation performance, i.e.,
as will be seen in Section~\ref{sec:CRB} lowering the CRB. The
transmit beamspace weight matrix ${\bf C}$ can be designed such
that the following two main requirement are satisfied: (i)~
Spatial distribution of energy transmitted within the desired
sector is uniform; (ii)~the amount of energy that is inevitably
transmitted in the out-of-sector area is minimized. Here we
present two methods for satisfying these two requirements.

\subsection{Spheroidal Sequences Based Transmit Beamspace Design}

Discrete prolate spheroidal sequences (DPSS) have been proposed
for beamspace dimension reduction in array processing
\cite{Vezzosi}.  The essence of the DPSS-based approach to
beamspace dimension reduction \cite{VanTrees}, \cite{Vezzosi} is
to maximize the ratio of the beamspace energy that comes from
within the desired sector ${\boldsymbol\Theta}$ to the total
beamspace energy, i.e., the energy within the whole spatial domain
$[-\pi/2,\ \pi/2]$. Following this principle, we propose to design
the transmit beamspace weight matrix so that the ratio of the
energy radiated within the desired spatial sector to the total
radiated energy is maximized. That is, the transmit beamspace
matrix ${\bf C}$ is designed based on maximizing
\begin{eqnarray}\label{eq:energy_ratio}
\Gamma_k &\!\!\!\triangleq\!\!\!&
\frac{\int_{T}\int_{\boldsymbol\Theta} \left|{\bf c}^H_k{\bf
a}(\theta)\phi_k(t)\right|^2 d\theta dt}
{\int_{T}\int_{-\frac{\pi}{2}}^{\frac{\pi}{2}} | {\bf c}^H_k {\bf
a}(\theta)\phi_k(t)|^2 \, d\theta dt} \nonumber\\
&\!\!\!=\!\!\!& \frac{{\bf
c}^H_k\left(\int_{\boldsymbol\Theta}{\bf a} (\theta){\bf
a}^H(\theta)d\theta\right){\bf
c}_k}{\int_{-\frac{\pi}{2}}^{\frac{\pi}{2}} |
{\bf c}^H_k {\bf a}(\theta)|^2 d\theta} \nonumber\\
&\!\!\!=\!\!\!&\frac{{\bf c}^H_k{\bf A}{\bf
c}_k}{\int_{-\frac{\pi}{2}}^{\frac{\pi}{2}} |{\bf c}^H_k {\bf
a}(\theta)|^2 d\theta}
\end{eqnarray}
where ${\bf A}\triangleq \int_{\boldsymbol\Theta}{\bf a}
(\theta){\bf a}^H(\theta)d\theta$ is a non-negative matrix. Note
that the first equality in \eqref{eq:energy_ratio} follows from
the fact that $\int_{T}\phi(t)\phi^{*}(t)dt = 1$ . Moreover, it
can be shown that if ${\bf a} (\theta)$ obeys Vandermonde
structure, then the following holds \cite{VanTrees}
\begin{equation}\label{eq:denominator}
\int_{-\frac{\pi}{2}}^{\frac{\pi}{2}} | {\bf c}^H_k {\bf
a}(\theta)|^2 d\theta = 2\pi {\bf c}^H_k{\bf c}_k.
\end{equation}
Substituting (\ref{eq:denominator}) in (\ref{eq:energy_ratio}), we
obtain
\begin{equation}\label{eq:energy_ratio1}
\Gamma_k = \frac{{\bf c}^H_k{\bf A}{\bf c}_k}{2\pi{\bf c}^H_k {\bf
c}_k},\quad k=1, \ldots, K.
\end{equation}
The maximization of the ration in \eqref{eq:energy_ratio1} is
equivalent to maximizing its numerator while fixing its
denominator by imposing the constraint $\|{\bf c}_k\|=1$. To avoid
the trivial solution ${\bf c}_1 = \ldots={\bf c}_K$, an additional
constraint might be necessary, for example, the orthogonality
constraint ${\bf c}^H_k{\bf c}_{k^{\prime}} = 0,\ k\neq
k^{\prime}$ can be imposed. Then, the maximization of
$\{\Gamma_k\}_{k=1}^{K}$ subject to the constraint ${\bf C}^H{\bf
C} = {\bf I}$ corresponds to finding the eigenvectors of ${\bf A}$
that are associated with the $K$ largest eigenvalues of ${\bf A}$.
That is, the transmit beamspace matrix is given as
\begin{equation}\label{Eq:ssbs}
{\bf C} = [{\bf u}_1, {\bf u}_2, \ldots, {\bf u}_{K}]
\end{equation}
where $\{{\bf u}_i\}_{i=1}^{K}$ are $K$ principal eigenvectors
of $\bf A$.

It is worth noting that any steering vector ${\bf a}(\theta),
\theta \in \boldsymbol \Theta$ belongs to the space spanned by the
columns of ${\bf C}$. This means that the magnitude of the
projector of ${\bf a}(\theta)$ onto the column space of ${\bf C}$
is approximately constant, i.e., the transmit power distribution
\begin{equation} \label{beampattern}
H (\theta) = {\bf a}^H ( \theta ) {\bf C} {\bf C}^H {\bf a}
\end{equation}
is approximately constant $\forall \theta \in \boldsymbol\Theta$.
Therefore, the total transmit power distribution, i.e., the
distribution of the power summed over all waveforms, within the
spatial sector $\boldsymbol \Theta$ is approximately uniform.
However, the transmit power distribution over individual waveforms
may not be uniform within $\boldsymbol \Theta$. Such a uniform
transmit power distribution over individual waveforms may be
desired especially when the difference between the first and the
last essential eigenvalues of ${\bf A}$ is significant. In this
respect, note that the orthogonality constraint ${\bf C}^H{\bf C}
= {\bf I}$ is imposed above only to avoid the trivial solution
${\bf c}_1 = \ldots={\bf c}_K$ when maximizing
\eqref{eq:energy_ratio1}. This is different from the case of the
traditional beamspace dimension reduction techniques where
orthogonality is required to preserve the white noise property.
Therefore, in our case where the latter is not an issue, ${\bf C}$
can be easily rotated, if necessary, so that it achieves desirable
features such as uniform transmit power distribution over
individual waveforms. The orthogonality may be lost after such
rotation of ${\bf C}$ without any consequences. For example, such
rotation of ${\bf C}$ is used in our simulations (Section~VII) to
ensure that the uniform transmit power distribution per each
individual waveform is achieved.

\subsection{Convex Optimization Based Transmit Beamspace Design}
A more general transmit beamspace design technique applicable to a
transmit array of arbitrary geometry can also be formulated.
Moreover, some DOA estimation techniques can be applied only if
the received data enjoy certain properties. For example, the
ESPRIT DOA estimation technique requires that the received data
snapshots consist of two sets where one data set is equivalent to
the other up to some phase rotation. This property is commonly
referred to as rotational invariance. More generally,
PARAFAC-based DOA estimation techniques are based on the
assumption that multiple received data sets are related via
rotational invariance properties. Fortunately, the transmit
beamspace matrix ${\bf C}$ can be designed so that the virtual
data vectors given by \eqref{eq:kthBeamComponent} enjoy such
rotational invariance property.

To obtain the rotational invariance property, we propose a convex
optimization based method for designing ${\bf C}$. For the virtual
snapshots $\{{\bf y}_{k}(\tau)\}^K_{k=1}$ to enjoy rotational
invariance, the following relationship should be satisfied
\begin{equation}\label{eq:Rotational}
e^{\jmath\mu_k(\theta)} \left| \left( {\bf c}^H_k {\bf
a}(\theta)\right) \right| {\bf b}(\theta) = e^{\jmath
\mu_{k^{\prime}} (\theta)} \left| \left( {\bf c}^H_{k^{\prime}}
{\bf a}(\theta) \right) \right| {\bf b}(\theta), \quad \forall
\theta \in \boldsymbol \Theta, \quad k, {k^{\prime}} = 1, \ldots,
K
\end{equation}
where $\jmath\triangleq\sqrt{-1}$, and $\mu_k(\theta)$ and
$\mu_{k^{\prime}}(\theta)$ are arbitrary phase shifts associated
with the $k$th and ${k^{\prime}}$th virtual snapshots,
respectively. The relationship \eqref{eq:Rotational} can be
satisfied if the transmit beamspace matrix ${\bf C}$ is
designed/optimized to meet the following requirement
\begin{equation}\label{eq:RotationalConstraint}
{\bf C}^H {\bf a}(\theta) = {\bf d}(\theta), \quad \forall \theta
\in \boldsymbol \Theta.
\end{equation}
where ${\bf d}(\theta)\triangleq [e^{\jmath\mu_1(\theta)}, \ldots,
e^{\jmath\mu_K(\theta)}]^T$ is of dimension $K\times 1$. A
meaningful formulation of a corresponding optimization problem is
to minimize the norm of the difference between the left- and
right-hand sides of \eqref{eq:RotationalConstraint} while keeping
the worst transmit power distribution in the out-of-sector area
below a certain level. This can be mathematically expressed as
follows
\begin{eqnarray}\label{eq:TransmitOptimization}
&&\min_{\bf C} \max_{i} \|{\bf C}^H{\bf a}({\theta}_i)-{\bf d}
(\theta)\| ,\quad {\theta}_i \in {\boldsymbol\Theta},\quad
i=1,\ldots,I \\
& &{\rm subject\ to}\quad \|{\bf C}^H{\bf a}({\theta}_j)\|
\leq \gamma , \quad {\theta}_j \in \bar{\boldsymbol\Theta},\
\ \ j=1,\ldots,J\label{eq:SidelobeLevels}
\end{eqnarray}
where $\bar{\boldsymbol\Theta}$ combines a continuum of all
out-of-sector directions, i.e., directions lying outside the
sector-of-interest $\boldsymbol\Theta$, and $\gamma>0$ is the
parameter of the user choice that characterizes the worst
acceptable level of transmit power radiation in the out-of-sector
region. The parameter $\gamma$ has an analogy to the stop-band
attenuation parameters in the classic bandpass filter design and
can be chosen in a similar fashion \cite{DigitalFilterDesign}.

\section{Transmit Beamspace Based DOA Estimation}
In this section, we design DOA estimation methods based on
transmit beamspace processing in MIMO radar. We focus on subspace
based DOA estimation techniques such as MUSIC and ESPRIT.

\subsection{Transmit Beamspace Based MUSIC}
The virtual data model \eqref{eq:virtual_data1} can be rewritten
as
\begin{eqnarray}\label{eq:virtual_data2}
{\bf y}_{\rm beam}(\tau) &\!\!\!=\!\!\!& {\bf V}{\boldsymbol\alpha}
(\tau) + {\tilde{\bf z}}_K(\tau)
\end{eqnarray}
where
\begin{eqnarray}\label{eq:TxbeamSterVec}
  {\boldsymbol\alpha}(\tau) &\!\!\!\triangleq\!\!\!&
  [\alpha_1(\tau),\ldots,\alpha_L(\tau)]^T \\
  {\bf V} &\!\!\!\triangleq\!\!\!& [{\bf v}(\theta_1),
  \ldots,{\bf v}(\theta_L)] \\
  {\bf v}(\theta) &\!\!\!\triangleq\!\!\!& \sqrt{\frac{E}{K}}
  \left({\bf C}^H{\bf a}(\theta)\right)\otimes {\bf b}(\theta)
  \label{eq:TxBeamSV}.
\end{eqnarray}

%%%%%%%%%%%%%%%%%%%
The $KN\times KN$ transmit beamspace-based covariance matrix is
given by
\begin{eqnarray}\label{eq:virtual_covariance}
{\bf R}_{\rm beam}&\!\!\!\triangleq\!\!\!& {\rm E}
\left\{{\bf y}_{\rm beam}(\tau){\bf y}_{\rm beam}^H(\tau)\right\}
\nonumber\\
&\!\!\!=\!\!\!& {\bf V}{\bf S}{\bf V}^H + \sigma_z^2{\bf I}_{KN}
\end{eqnarray}
where ${\bf S}\triangleq{\rm E}\{ {\boldsymbol \alpha} (\tau)
{\boldsymbol \alpha}^H (\tau) \}$ is the covariance matrix of the
reflection coefficients vector. The sample estimate of
(\ref{eq:virtual_covariance}) takes the following form
\begin{equation}\label{SampleCovMatrTxBeam}
\hat{\bf R}_{\rm beam} = \frac{1}{Q} \sum_{\tau=1}^Q {\bf y}_{\rm
beam} (\tau){\bf y}^H_{\rm beam}(\tau).
\end{equation}

The eigendecomposition of (\ref{SampleCovMatrTxBeam}) can be
written as
\begin{equation}
\hat{\bf R}_{\rm beam} = {\bf E}_{\rm s}{\boldsymbol \Lambda}_{\rm
s}{\bf E}_{\rm s}^H + {\bf E}_{\rm n}{\boldsymbol \Lambda}_{\rm
n}{\bf E}_{\rm n}^H \label{eigendec}
\end{equation}
where the $L\times L$ diagonal matrix ${\boldsymbol \Lambda}_{\rm
s}$ contains the largest (signal-subspace) eigenvalues and the
columns of the $KN\times L$ matrix ${\bf E}_{\rm s}$ are the
corresponding eigenvectors. Similarly, the $(KN-L)\times(KN-L)$
diagonal matrix ${\boldsymbol \Lambda}_{\rm n}$ contains the
smallest (noise-subspace) eigenvalues while the $KN\times(KN-L)$
matrix ${\bf E}_{\rm n}$ is built from the corresponding
eigenvectors.

Applying the principle of the elementspace MUSIC estimator
\cite{Schmidt2}, the transmit beamspace spectral-MUSIC estimator
can be expressed as
\begin{equation}\label{TxbsMUSIC}
f(\theta) = \frac{{\bf v}^H(\theta){\bf v}(\theta)}{{\bf v}^H
(\theta){\bf Q}{\bf v}(\theta)}
\end{equation}
where ${\bf Q}={\bf E}_{\rm n}{\bf E}_{\rm n}^H={\bf I} - {\bf
E}_{\rm s}{\bf E}_{\rm s}^H$ is the projection matrix onto the
noise subspace. Substituting \eqref{eq:TxBeamSV} into
\eqref{TxbsMUSIC}, we obtain
\begin{eqnarray}\label{TxbsMUSIC1}
f(\theta)&\!\!\!=\!\!\!& \frac{\left[\left({\bf C}^H{\bf a}
(\theta)\right) \otimes {\bf b}(\theta)\right]^H  \left[\left(
{\bf C}^H{\bf a}(\theta)\right)\otimes {\bf b}(\theta)\right]}{\left[
\left({\bf C}^H{\bf a}(\theta)\right)\otimes {\bf b}(\theta)\right]^H
{\bf Q} \left[\left({\bf C}^H{\bf a}(\theta)\right)\otimes {\bf b}
(\theta)\right]}\nonumber\\
&\!\!\!=\!\!\!&\frac{[{\bf a}^H(\theta){\bf C}{\bf C}^H{\bf a}
(\theta)] \cdot[{\bf b}^H(\theta){\bf b}(\theta)]}
{\left[\left({\bf C}^H{\bf a} (\theta)\right)\otimes {\bf
b}(\theta)\right]^H {\bf Q} \left[\left(
{\bf C}^H {\bf a} (\theta) \right)\otimes {\bf b}(\theta)\right]}
\nonumber\\
&\!\!\!=\!\!\!&\frac{N{\bf a}^H(\theta){\bf C}{\bf C}^H{\bf a}
(\theta)} {\left[\left({\bf C}^H{\bf a}(\theta)\right)\otimes {\bf
b} (\theta) \right]^H {\bf Q} \left[\left({\bf C}^H{\bf
a}(\theta)\right)\otimes {\bf b}(\theta) \right]}.
\end{eqnarray}

\subsection{Transmit Beamspace Based ESPRIT}
The signal component of the data vectors $\{{\bf
y}_{k}(\tau)\}_{k=1}^K$ can be expressed as
\begin{equation}\label{eq:data_vectors}
{\bf y}_{k}(\tau) = {\bf T}_k{\boldsymbol\alpha}(\tau)
\end{equation}
where
\begin{eqnarray}
{\bf T}_k &\!\!\!\triangleq\!\!\!&[{\bf b}(\theta_1),\ldots,{\bf
b}(\theta_L)]{\boldsymbol\Psi}_k \\
{\boldsymbol\Psi}_k&\!\!\!\triangleq\!\!\!&{\rm diag}\left\{{\bf
c}_k^H {\bf a}(\theta_1),\ldots,{\bf c}_k^H{\bf
a}(\theta_L)\right\}.\label{eq:rotation_matrix}
\end{eqnarray}

It is worth noting that the matrices $\{{\bf T}_k\}_{k=1}^K$ are
related to each other as
\begin{equation}\label{eq:data_rotation}
{\bf T}_k = {\bf T}_j{\boldsymbol\Psi}_j^{-1}{\boldsymbol\Psi}_k,
\quad k,j = 1,\ldots,K.
\end{equation}
By carefully designing the beamspace weight matrix ${\bf C}$, for
example by using
\eqref{eq:TransmitOptimization}--\eqref{eq:SidelobeLevels}, the
relationship (\ref{eq:data_rotation}) enjoys the rotational
invariance property.

Consider the case when only two transmit beams are formed. Then,
the transmit beamspace matrix is ${\bf C}=[{\bf c}_1,{\bf c}_2]$.
In this case, (\ref{eq:data_rotation}) simplifies to
\begin{equation}\label{eq:rotation_invariance}
{\bf T}_2 = {\bf T}_1{\boldsymbol\Psi}
\end{equation}
where
\begin{eqnarray}\label{eq:rotations}
{\boldsymbol\Psi}&\!\triangleq\!& {\rm diag}\left\{\frac{{\bf
c}_2^{H} {\bf a} (\theta_1)} {{\bf c}_1^{H}{\bf a} (\theta_1)},
\ldots, \frac{{\bf c}_2^{H}{\bf a} (\theta_L)}{{\bf c}_1^{H} {\bf
a} (\theta_L)}\right\} .
\end{eqnarray}
Furthermore, (\ref{eq:rotations}) can be rewritten as
\begin{eqnarray}\label{eq:rotations1}
{\boldsymbol\Psi}&\!=\!& {\rm diag} \left\{A(\theta_1)e^{\jmath
\Omega(\theta_1)},\ldots, A(\theta_L) e^{\jmath \Omega(\theta_L)}
\right\}
\end{eqnarray}
where $A(\theta)$ and $\Omega(\theta)$ are the magnitude and angle
of ${\bf c}_2^{H}{\bf a} (\theta) / {\bf c}_1^{H}{\bf a}
(\theta)$, respectively. It can be seen from (\ref
{eq:rotation_invariance})~and~(\ref{eq:rotations1}) that the
vectors ${\bf y}_1$ and ${\bf y}_2$ (see also
(\ref{eq:data_vectors})) enjoy the rotational invariance property.
Therefore, ESPRIT-based DOA estimation techniques can be used to
estimate ${\boldsymbol\Psi}$ and $\{\theta_l\}_{l=1}^L$ can be
obtained from ${\boldsymbol\Psi}$ by looking up a table that
converts $\Omega(\theta)$ to $\theta$. Moreover, if $K>2$ is
chosen, more than two data sets which enjoy the rotational
invariance property can be obtained by properly designing the
transmit beamspace weight matrix. In this case, the means of using
PARAFAC instead of ESPRIT are provided as well.

It is worth mentioning that for traditional MIMO radar
(\ref{eq:virtual_data}), DOA estimation using ESPRIT has been
proposed in \cite{ESPRIT_MIMO_radar_2008}. Specifically, ${\bf
y}(\tau)$ in (\ref{eq:virtual_data}) has been partitioned into
${\bf y}_1(\tau)~\triangleq~[{\bf x}^T_{1}(\tau),$ $\ldots,{\bf
x}^T_{M-1}(\tau)]^T$ and ${\bf y}_2(\tau)~\triangleq~[{\bf
x}^T_{2}(\tau),$ $\ldots,{\bf x}^T_{M}(\tau)]^T$ and it has been
shown that ${\bf y}_1$ and ${\bf y}_2$ obey the rotational
invariance property which enables the use of ESPRIT for DOA
estimation. However, the rotational invariance property is valid
in \cite{ESPRIT_MIMO_radar_2008} only when the transmit array is a
uniform linear array (ULA). Therefore, the method of
\cite{ESPRIT_MIMO_radar_2008} is limited by the transmit array
structure and may suffer from performance degradation in the
presence of array perturbation errors. Moreover, it suffers from
low SNR per virtual antenna as a result of dividing the total
transmit energy over $M$ different waveforms.

\section{Performance Analysis}
In this section, we analyze the performance of the proposed
transmit beamspace MIMO radar DOA estimation approach as compared
to the MIMO radar DOA estimation technique. For the reason of
comparison, we also consider two techniques that have been
recently reported in the literature that employ the idea of
dividing the transmit array into several smaller
subapertures/subarrays. The first technique uses transmit
subapertures (TS) for omni-directionally radiating independent
waveforms \cite{RabideauParker}, \cite{Bergin2008}. If the number
of subapertures is chosen as $K<M$, then each subaperture radiates
pulses of energy $E/K$. The second technique is based on
partitioning the transmit array into overlapped subarrays where
the antennas that belong to each subarray are used to coherently
transmit an independent waveform \cite{NasrSergConf},
\cite{NasrSerg}. We refer to this technique as transmit array
partitioning (TAP). Note that the TAP technique has transmit
coherent gain while the TS technique does not have such a coherent
transmit gain.

In the following subsection, we compare between the aforementioned
techniques in terms of the effective aperture of the corresponding
virtual array, the SNR gain per virtual element, and the
computational complexity associated with eigendecomposition based
DOA estimation techniques. In the next subsection, we
express/discuss the CRB for all considered techniques.

\subsection{Colocated Uniform Linear Transmit Array}
Consider the case of a ULA at the transmitter with $\lambda/2$
spacing between adjacent antennas, where $\lambda$ is the
propagation wavelength. Taking the first antenna as a reference,
the transmit steering vector can be expressed as
\begin{equation}\label{eq:transmitSV}
{\bf a}(\theta)\triangleq\left[1, e^{-j \pi \sin\theta}, \ldots,
e^{-j \pi (M-1)\sin\theta}\right]^T.
\end{equation}
Also, the receive antennas are assumed to be grouped in a ULA with
half a wavelength interelement spacing. Then, the receive steering
vector is given as
\begin{equation}\label{eq:receiveSV}
{\bf b}(\theta)\triangleq\left[1, e^{-j \pi \sin\theta}, \ldots,
e^{-j \pi (N-1)\sin\theta}\right]^T.
\end{equation}
It is worth noting that the transmit and receive array apertures
are $(M-1)\lambda/2$ and $(N-1)\lambda/2$, respectively. In light
of \eqref{eq:transmitSV} and \eqref{eq:receiveSV}, we
discuss/analyze the following cases.

\subsubsection{Traditional MIMO radar}
Substituting \eqref{eq:transmitSV} and \eqref{eq:receiveSV} in
\eqref{eq:MIMOvirtualSV}, the MIMO radar virtual steering vector
can be expressed as
\begin{eqnarray}\label{eq:virtualSV}
{\bf u}_{\rm MIMO}(\theta)&\!\!\!=\!\!\!&\left[1, \ldots,
e^{-j\pi(N-1)\sin\theta}, e^{-j\pi\sin\theta}, \ldots,
e^{-j\pi N\sin\theta},\right.\nonumber\\
&&\left. \ldots, e^{-j\pi(M-1)\sin\theta},\ldots, e^{-j 2\pi (M-1
+ N-1)\sin\theta} \right]^T.
\end{eqnarray}
From \eqref{eq:virtualSV}, we observe that the effective virtual
array aperture is $(M + N -2)\lambda/2$. Note that the virtual
steering vector ${\bf a}(\theta)\otimes {\bf a}(\theta)$ is of
dimension $MN \times 1$, yet it only contains $M+N-1$ distinct
elements. Moreover, the SNR gain per virtual element is
proportional  in this case to $E/M$. This low SNR gain can lead to
poor DOA estimation performance especially at low SNR region. The
computational complexity of applying eigen-decomposition  based
DOA estimation techniques is of $O(M^3N^3)$ in this case.

\subsubsection{Transmit subaperturing based MIMO radar}
As compared to the traditional MIMO radar, TS-based MIMO radar
employs $K$ subapertures instead of $M$. This results in higher
SNR per virtual element in the corresponding virtual array at the
receiver. To capitalize on the effect of this factor on the DOA
estimation performance, we consider the extreme case when only two
transmit subapertures\footnote{One can think of a subaperture as a
large omni-directional antenna which is capable of radiating
energy $E/2$ instead of $E/M$. This case might be practically
unattractive as it will require power amplifier of much higher
amplifying gain as compared to the case of using $M$ transmit
antennas. However, for the sake of theoretical analysis/comparison
we consider it in this paper.} are used to radiate the total
transmit energy $E$. In this case, each transmit subaperture
radiates a waveform of energy $E/2$. Assume that the two transmit
subapertures are separated in space by $\zeta$ wavelength and that
the first transmit subaperture is taken as a reference. Then, the
MIMO radar data model \eqref{eq:virtual_data} becomes of dimension
$2N\times 1$ and can be expressed as
\begin{eqnarray}\label{eq:virtual_twoTx}
{\bf y}_{\rm TS} (\tau) &\!\!\!=\!\!\!&
\sqrt{\frac{E}{2}}\sum_{l=1}^{L} \alpha_l(\tau)\left([1,\  e^{-j
2\pi \zeta \sin \theta_l}]^T \otimes {\bf b} (\theta) \right) +
{\tilde{\bf z}}_{\rm TS}(\tau)
\end{eqnarray}
where ${\tilde{\bf z}}_{\rm TS}(\tau) = [{\bf z}_1 (\tau), \ {\bf
z}_2 (\tau)]^T$.

Comparing \eqref{eq:virtual_data} to \eqref{eq:virtual_twoTx}, we
observe that the signal strength for MIMO radar with $M$ transmit
antennas is proportional to $\sqrt{E/M}$ while the signal strength
for the TS-based MIMO radar is proportional to $\sqrt{E/2}$. This
means that \eqref{eq:virtual_twoTx} offers  an SNR gain that is
$M/2$ times the SNR gain offered by \eqref{eq:virtual_data}. We
also observe from \eqref{eq:virtual_twoTx} that the effective
virtual array aperture is given by $(\zeta + N-1)\lambda/2$.
Therefore, the effective aperture for this case can be controlled
by selecting $\zeta$. For example, selecting $\zeta=\lambda/2$
yields a virtual array steering vector of dimension $2N$ that
contains only $N+1$ distinct elements, i.e, the effective aperture
would be $N\lambda/2$. This case is particularly important when
performing DOA estimation using search free techniques such as
ESPRIT. Another important case is the choice $\zeta=N\lambda/2$
which yields a virtual array that is equivalent to a $2N$-element
ULA. In this case, the effective array aperture will be
$(2N-1)\lambda/2$. The computational complexity of applying
eigen-decomposition  based DOA estimation techniques is of
$O(2^3N^3)$ in this case.

\subsubsection{Transmit array partitioning based MIMO radar}
Following the guidelines of \cite{NasrSerg} and selecting $K=2$,
the $M$-antenna transmit array is assumed to be partitioned into
two fully overlapped subarrays of $M-1$ antenna each. The
$(M-1)\times 1$ beamforming weight vectors ${\bf w}_1 = {\bf
w}_2={\bf w}$ are used to form transmit beams that cover the
spatial sector $\boldsymbol\Theta$. Two independent waveforms are
radiated. Each waveform has $E/2$ energy per pulse. The transmit
weight vector ${\bf w}$ can be designed such that the transmit
gain is approximately the same within the desired sector
$\boldsymbol \Theta$, i.e., $|{\bf w}^H{\bar{\bf a}(\theta)} | =
G_{\rm TAP},\ \forall \theta \in \boldsymbol \Theta$ where
${\bar{\bf a}(\theta)}$ contains the first $M-1$ elements of the
vector ${\bf a}(\theta)$ and $G_{\rm TAP}$ is the TAP transmit
coherent processing gain. Then, the TAP-based MIMO radar data
model becomes of dimension $2N\times 1$ and can be formally
expressed as
\begin{eqnarray}\label{eq:virtual_TAP}
{\bf y}_{\rm TAP}(\tau) &\!\!\!=\!\!\!& \sqrt{\frac{E}{2}}
\sum_{l=1}^{L} \alpha_l (\tau) \left( {\bf w}^H {\bar{\bf
a}(\theta_l)} \right) {\bf u}_{\rm TAP} (\theta_l) + {\tilde{\bf
z}}_{\rm TAP} (\tau)
\nonumber\\
&\!\!\!=\!\!\!& G_{\rm TAP} \sqrt{\frac{E}{2}} \sum_{l=1}^{L}
\alpha_l (\tau) {\bf u}_{\rm TAP} (\theta_l) + {\tilde{\bf
z}}_{\rm TAP} (\tau)
\end{eqnarray}
where ${\bf u}_{\rm TAP} (\theta) \triangleq [1,\ e^{-j \pi \sin
\theta}]^T \otimes {\bf b} (\theta)$ is the $2N\times 1$ steering
vector of the corresponding virtual array. Note that the TAP-based
MIMO radar has transmit coherent gain $G_{\rm TAP}$ which results
in improvement in SNR per virtual element. However, the
corresponding virtual array contains $N+1$ distinct elements,
i.e., the effective virtual array aperture is limited to
$N\lambda/2$. The computational complexity of applying
eigen-decomposition  based DOA estimation techniques is of
$O(2^3N^3)$ in this case.

\subsubsection{Transmit beamspace based MIMO radar}
For the proposed transmit beamspace MIMO radar, we choose $K=2$
and use \eqref{eq:TransmitOptimization}--\eqref{eq:SidelobeLevels}
for designing ${\bf C}=[{\bf c}_1\ {\bf c}_2]$ such that
\begin{equation}\label{eq:transmitBeamTwo}
{\bf d}(\theta) = G_{\rm beam} \cdot [1,\ e^{-j 2\pi
N\sin\theta}]^T, \quad \forall\theta\in\boldsymbol\Theta
\end{equation}
where $G_{\rm beam}$ is the transmit beamspace gain, i.e., $| {\bf
c}^H_1 {\bf a} (\theta) | \approx | {\bf c}^H_2 {\bf a} (\theta) |
\approx G_{\rm beam}, \ \forall \theta \in \boldsymbol \Theta$.
This yields a virtual array with $2N$ distinct elements and
$(2N-1)\lambda/2$ effective array aperture. Moreover, the proposed
transmit beamspace technique offers SNR gain of $G_{\rm beam}
\cdot E / M$, i.e., it combines all the benefits of all other
aforementioned techniques. The computational complexity of
applying eigen-decomposition based DOA estimation techniques is of
$O(2^3N^3)$ in this case.

A comparison between all methods considered is summarized in the
following table. \\

Table1: Comparison between transmit beamspace-based MIMO radar and
other existing techniques. \vspace{-.5cm}
\begin{table}[h!]
  \centering
\begin{tabular}{|c|c|c|c|}
  \hline
  \ & Effective aperture & SNR gain per virtual element & Computational complexity  \\
  \hline
  Traditional MIMO (5)\hfill\ & $(M+N-2) \frac{\lambda}{2}$ & $\frac{E}{M}$ &
  $O(M^3N^3)$ \\
  Transmit subaperturing ($\zeta= \frac{\lambda}{2}$)\hfill\  & $N \frac{\lambda}{2}$
  & $\frac{E}{2}$ & $O(2^3N^3)$  \\
  Transmit subaperturing ($\zeta = N \frac{\lambda}{2}$)\hfill\  & $(2N-1)
  \frac{\lambda}{2}$ & $\frac{E}{2}$ & $O(2^3N^3)$   \\
  Transmit array partitioning \eqref{eq:virtual_TAP}\hfill\  & $N \frac{\lambda}{2}$ &
  $G^2_{\rm TAP} \cdot \frac{E}{2}$ & $O(2^3N^3)$ \\
  Transmit beamspace MIMO \eqref{eq:transmitBeamTwo}\hfill\  & $(2N-1)
  \frac{\lambda}{2}$ & $G^2_{\rm beam} \cdot \frac{E}{2}$  & $O(2^3N^3)$ \\
  \hline
\end{tabular}
\label{Table1}
\end{table}

\subsection{Cram\'er-Rao Bound}\label{sec:CRB}
In this section, we discuss the CRB on DOA estimation accuracy in
transmit beamspace-based MIMO radar.

In the case of transmit beamspace-based MIMO radar, the virtual
data model \eqref{eq:virtual_data1} satisfies the following
statistical model:
\begin{equation}
{\bf y}_{\rm beam}(\tau)\sim {\cal N}_C \left( {\boldsymbol
\mu}(\tau), {\bf R}\right) \label{statmod}
\end{equation}
where ${\cal N}_C$ denotes the complex multivariate circularly
Gaussian probability density function, ${\boldsymbol\mu}(\tau)$ is
the mean of ${\bf y}_{\rm beam}(\tau)$, and ${\bf R}$ is its
covariance matrix.

\subsubsection{Stochastic CRB} The stochastic CRB on estimating
the DOAs using the data model \eqref{eq:virtual_data1} is derived
by assuming ${\boldsymbol\mu}(\tau)={\bf 0}$ and ${\bf R} = {\bf
R}_{\rm beam}$, where ${\bf R}_{\rm beam}$ is given by
\eqref{eq:virtual_covariance}. Note that under these assumptions,
the virtual array signal model \eqref{eq:virtual_data1} (or its
equivalent representation \eqref{eq:virtual_data2}) has the same
form as the signal model used in \cite{StoicaLarGersh} to derive
the stochastic CRB for DOA estimation in conventional array
processing. Therefore, the CRB expressions in its general form
derived in \cite{StoicaLarGersh} can be used for computing the
stochastic CRB for estimating the DOAs based on
\eqref{eq:virtual_data2}, that is,
\begin{equation}\label{eq:CRB}
{\rm CRB}(\theta) = \frac{\sigma_z^2}{2Q}\left\{{\rm Re}({\bf D}
{\bf P}^{\bot}_{\bf V}{\bf D})\odot{\bf G}^T\right\}^{-1}
\end{equation}
where ${\bf P}^{\bot}_{\bf V}\triangleq{\bf V}({\bf V}^H{\bf
V})^{-1}{\bf V}^H$ is the projection matrix onto the space spanned
by the columns of ${\bf V}$, and ${\bf G}\triangleq({\bf S}{\bf
V}^H{\bf R}^{-1}{\bf V}{\bf S})$. In \eqref{eq:CRB}, ${\bf
D}\triangleq[{\bf d}(\theta_1),\ldots,{\bf d}(\theta_L)]$ is the
matrix whose $l$th column is given by the derivative of the $l$th
column of ${\bf V}$ with respect to $\theta_l$, i.e.,
\begin{eqnarray}\label{eq:vector_derivative}
{\bf d}(\theta)&\!\!\!\triangleq\!\!\!&\frac{d{\bf v}(\theta)}{d\theta}
= \sqrt{\frac{E}{K}}\frac{d\left[({\bf C}^H{\bf a}(\theta))
\otimes {\bf b}(\theta)\right]}{d\theta}\nonumber\\
&\!\!\!=\!\!\!&\sqrt{\frac{E}{K}}\left( ({\bf C}^H{\bf a}^{\prime}
(\theta) ) \otimes {\bf b}(\theta) + ( {\bf C}^H {\bf a} (\theta)
) \otimes {\bf b}^{\prime} (\theta)\right)
\end{eqnarray}
where ${\bf a}^{\prime}(\theta)=d{\bf a}(\theta)/d\theta$ and
${\bf b}^{\prime}(\theta)=d{\bf b}(\theta)/d\theta$.

\subsubsection{Deterministic CRB}
The deterministic CRB is derived by assuming
${\boldsymbol\mu}(\tau) = {\bf V}{\boldsymbol\alpha}(\tau)$ and
${\bf R}=\sigma_z^2{\bf I}_{KN}$. Under these statistical
assumptions, the virtual data model \eqref{eq:virtual_data1} is
similar to the general model used in \cite{StoicaNehorai} to find
the deterministic CRB. According to \cite{StoicaNehorai}, the
deterministic CRB expression can be obtained from \eqref{eq:CRB}
by replacing ${\bf G}$ with ${\hat{\bf S}}$, where ${\hat{\bf S}}$
is the sample estimate of ${\bf S}$.

It is worth noting that the expression \eqref{eq:CRB} can be used
not only for computing the CRB for the proposed transmit beamspace
technique but also for the other techniques summarized in Table~1.
Indeed, the following cases show how these techniques can be
viewed as special cases of the proposed model
\eqref{eq:virtual_data1}.
\begin{enumerate}
  \item Choosing ${\bf C}={\bf I}_M$, the transmit beamspace signal
  model \eqref{eq:virtual_data1} simplifies to the traditional MIMO
  radar signal model \eqref{eq:virtual_data}. Therefore, the CRB for
  the traditional MIMO radar  can be obtained by substituting
  ${\bf C} = {\bf I}_M$ in \eqref{eq:CRB} and \eqref{eq:vector_derivative}.
  \item The TS-based MIMO radar signal model with $\zeta = \lambda/2$ can
  be obtained from \eqref{eq:virtual_data1} by choosing ${\bf C}$ in
  the following format
  \begin{equation}\label{eq:TwoTxAntennas1}
    {\bf C} = \begin{bmatrix}
                1 & 0 & {\bf 0} \\
                0 & 1 & {\bf 0} \\
              \end{bmatrix}^T.
  \end{equation}
  \item The TS-based MIMO radar signal modal for $\zeta = N \lambda/2$ can
  be obtained from \eqref{eq:virtual_data1} by choosing ${\bf C}$ in the
  following format
  \begin{equation}\label{eq:TwoTxAntennas2}
    {\bf C} = \begin{bmatrix}
                1 & {\bf 0} & 0 \\
                0 & {\bf 0} & 1 \\
              \end{bmatrix}^T.
  \end{equation}
  \item Finally, the TAP-based MIMO radar signal model \eqref{eq:virtual_TAP}
  can be obtained from \eqref{eq:virtual_data1} by choosing ${\bf C}$ in the
  following format
        \begin{equation}\label{eq:C_TAP}
    {\bf C} = \begin{bmatrix}
              {\bf w}_1  & 0 \\
                0 & {\bf w}_2 \\
              \end{bmatrix}.
\end{equation}
\end{enumerate}

\section{Simulation Results}\label{sec:simresults}
Throughout our simulations, we assume a uniform linear transmit
array of $M\!=\!10$ omni-directional antennas spaced half a
wavelength apart. At the receiver, $N\!=\!10$ omni-directional
antennas are also assumed. The additive noise is Gaussian
zero-mean unit-variance spatially and temporally white. Two
targets are located at directions $-1^{\circ}$ and $1^{\circ}$,
respectively. The sector of interest $\boldsymbol\Theta =
[-5^{\circ},\, 5^{\circ}]$ is taken. Several examples are used to
compare the performances of the following methods: (i)~The
traditional MIMO radar \eqref{eq:virtual_data}; (ii)~The TS-based
MIMO radar \eqref{eq:TwoTxAntennas1} with $\zeta=\lambda/2$;
(iii)~The TS-based MIMO radar \eqref{eq:TwoTxAntennas1} with
$\zeta=N\lambda/2$; (iv)~the TAP-based MIMO radar
\eqref{eq:C_TAP}; and (v)~The proposed transmit beamspace MIMO
radar \eqref{eq:virtual_data1}. For all methods tested, the total
transmit energy is fixed to $E=M$. For the traditional MIMO radar
\eqref{eq:virtual_data}, each transmit antenna is used for
omni-directional radiation of one of the baseband waveforms
\begin{equation}\label{eq:SimulatedWaveforms}
\phi_m (t) = \sqrt{\frac{1}{T}} e^{\jmath 2 \pi \frac{m}{T}
t},\quad {m=1,\ldots,M}.
\end{equation}
For all methods that radiate two waveforms only, the first and
second waveforms of \eqref{eq:SimulatedWaveforms} are used. The
total number of virtual snapshots used to compute the sample
covariance matrix is $Q = 300$. In all examples, the RMSEs and the
probability of source resolution for all methods tested are
computed based on $500$ independent runs.

\subsection{Example~1: Stochatstic and Deterministic CRBs}
\begin{figure}[ht]
\begin{minipage}[b]{1.0\linewidth}
  \centering
  \centerline{\epsfig{figure=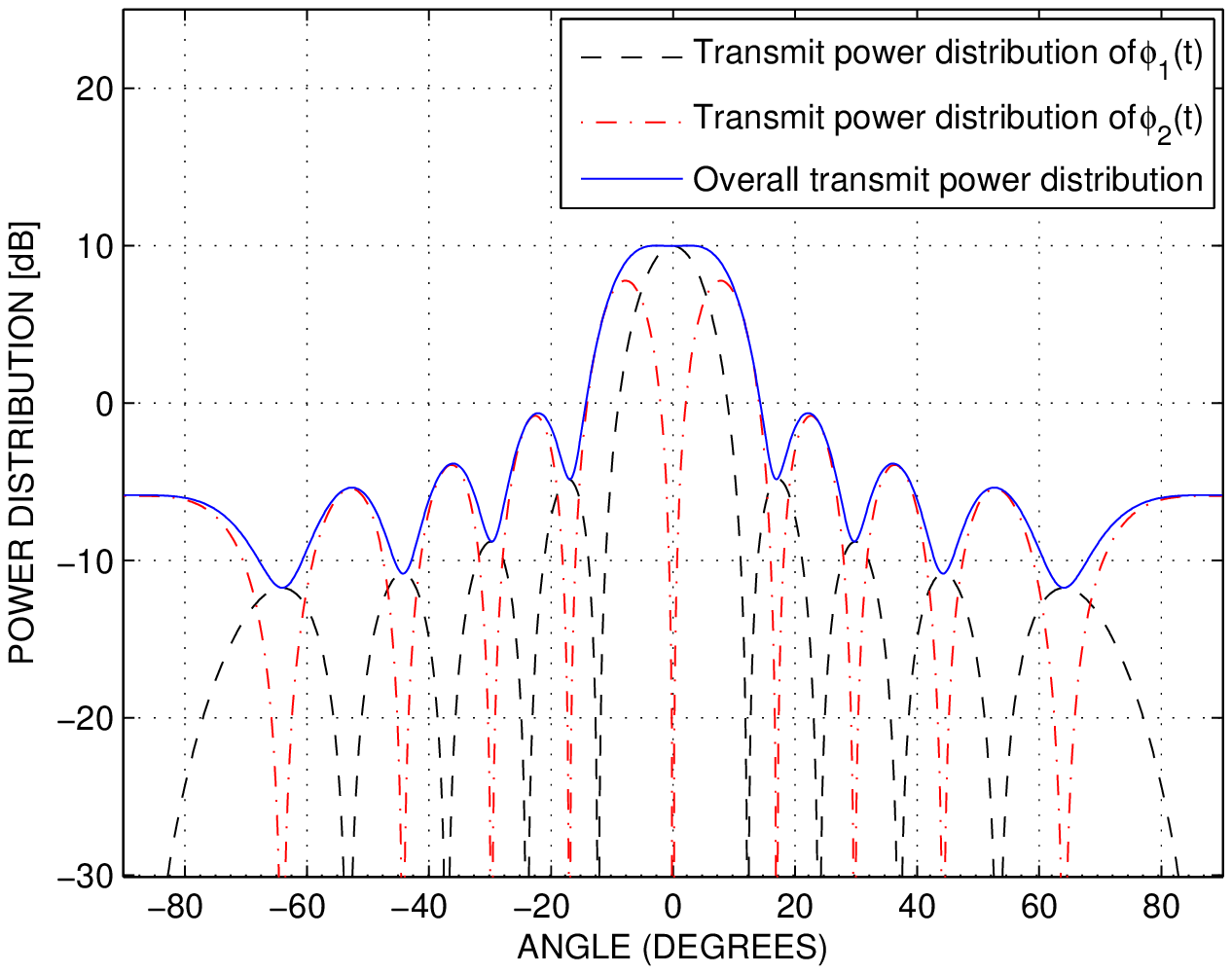,width=10.5cm}}
\end{minipage}
\caption{Transmit beamspace beampattern using spheroidal sequences
without rotation. Total transmit power is uniformly distributed
within the desired spatial sector, however, different transmitted
waveforms have different power distribution.} \label{fig:pattern1}
\end{figure}

\begin{figure}[ht]
\begin{minipage}[b]{1.0\linewidth}
  \centering
  \centerline{\epsfig{figure=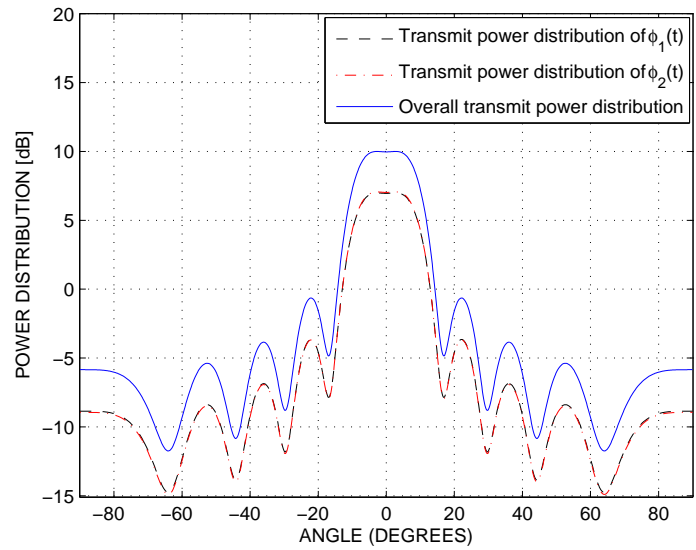,width=10.5cm}}
\end{minipage}
\caption{Transmit beamspace beampattern using spheroidal sequences
with rotation. Total transmit power as well as individual waveform
powers are uniformly distributed within the desired spatial
sector.} \label{fig:pattern2}
\end{figure}

For the proposed spheroidal sequences based transmit beamspace
MIMO radar \eqref{Eq:ssbs}, the non-negative matrix ${\bf A} =
\int_{\boldsymbol\Theta}{\bf a}(\theta){\bf a}^H(\theta)d\theta$
is built. The two eigenvectors associated with the largest two
eigenvalues are taken as the principle eigenvectors, i.e., ${\bf
C} = [{\bf u}_1 \ {\bf u}_2]$ is used. Two waveforms are assumed
to be radiated where each column of ${\bf C}$ is used to form a
transmit beam for radiating a single waveform.
Fig.~\ref{fig:pattern1} shows the transmit power distribution of
the individual waveforms $| {\bf c}_k {\bf a} ( \theta ) |^2$,
$k=1,2$ as well as the distribution of the total transmitted power
\eqref{beampattern}. As we can see from this figure, the
individual waveform power is not uniformly distributed within the
sector $\boldsymbol\Theta$ while the distribution of the total
transmitted power is uniform. To achieve uniform transmit power
distribution for each waveform a simple modification that involves
vector rotation to the transmit beamspace matrix ${\bf C}$ can be
performed. This means that ${\bf C}$ can be modified as
\begin{equation} \label{eq:ModifiedC}
{\tilde{\bf C}} = {\bf C}{\bf Q}
\end{equation}
where ${\bf Q}$ is a $2\times 2$ unitary matrix defined as
\begin{equation} \label{eq:UnitaryMatrix}
{\bf Q} = \begin{bmatrix}
            \sqrt{1/2} & \sqrt{1/2} \\
            \sqrt{1/2} & -\sqrt{1/2} \\
          \end{bmatrix}.
\end{equation}
Fig.~\ref{fig:pattern2} shows the transmit power distribution for
the individual waveforms $| \tilde{{\bf c}}_k {\bf a} ( \theta )
|^2$, $k=1,2$ as well as for the total transmitted power
\eqref{beampattern}. It can be seen from this figure that the
distribution of individual waveforms is uniform within the desired
sector.

\begin{figure}[ht]
\begin{minipage}[b]{1.0\linewidth}
  \centering
  \centerline{\epsfig{figure=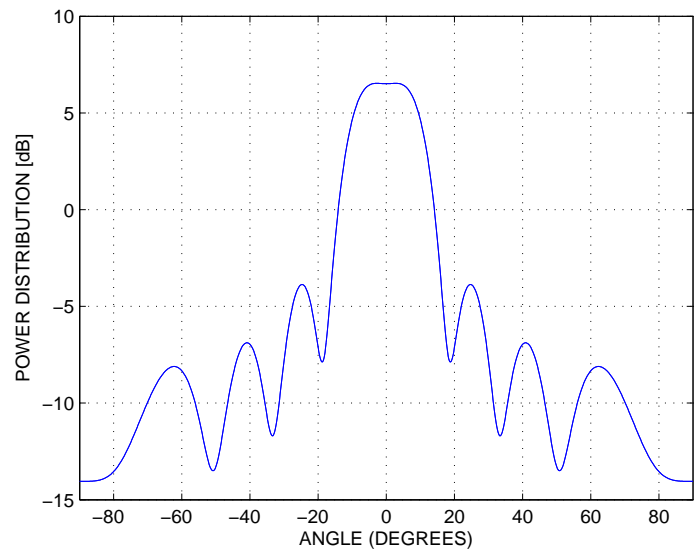,width=10.5cm}}
\end{minipage}
\caption{TAP-based MIMO radar transmit beampattern using two fully
overlapped subarrays.} \label{fig:pattern}
\end{figure}

For the TAP-based method, the transmit array is partitioned into
two overlapped subarrays of $9$ antennas each. Each subarray is
used to focus the radiation of one waveform within the sector
$\boldsymbol\Theta$. The transmit weight vectors are chosen as
${\bf w}_1 = {\bf w}_2 = [-0.5623\,$ $-0.5076\,$ $-0.4358\,$
$-0.3501\,$ $-0.2542\,$ $-0.1524\,$ $-0.0490\,$ $0.0512\,$
$0.1441]^T$. This specific selection is obtained by averaging the
two eigenvectors associated with the maximum two eignevalues of
the matrix $\int_{\boldsymbol\Theta}{\bf a}_1(\theta){\bf
a}^H_1(\theta)d\theta$, where ${\bf a}_1(\theta)$ is the $9\times
1$ steering vector associated with the first
subarray.\footnote{Note that if the transmit array is not a ULA,
then ${\bf w}_1$ and ${\bf w}_2$ can be designed independently
using classic FIR filter design techniques.} The transmit power
distribution of both waveforms is exactly the same as shown in
Fig.~\ref{fig:pattern}.

\begin{figure}[ht]
\begin{minipage}[b]{1.0\linewidth}
  \centering
  \centerline{\epsfig{figure=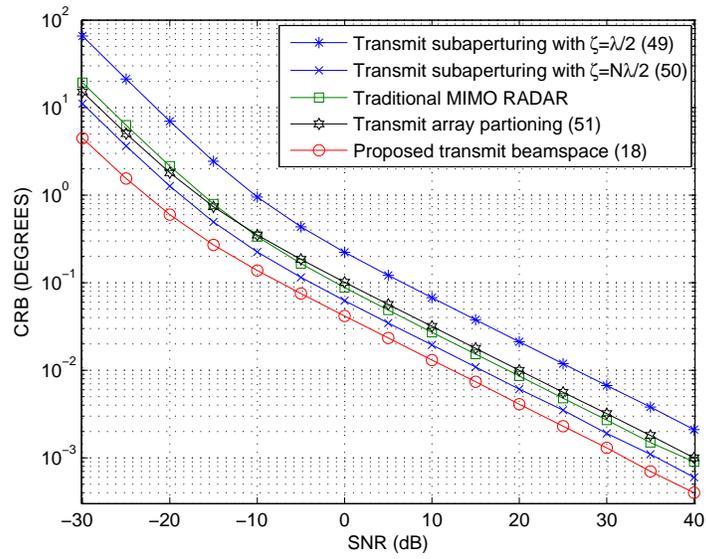,width=10.5cm}}
\end{minipage}
\caption{Stochastic CRB versus SNR.} \label{fig:StoCRB}
\end{figure}

The stochastic CRBs for all methods considered are plotted versus
${\rm SNR} = {\sigma_{\alpha}^2}/ {\sigma_z^2}$ in
Fig.~\ref{fig:StoCRB}. It can be seen from this figure that the
TS-based MIMO radar with $\zeta=\lambda/2$ has the highest/worst
CRB as compared to all other methods. Its poor CRB performance is
attributed to the omni-directional transmission, i.e., wasting a
considerable fraction of the transmitted energy within the
out-of-sector region, and to the small effective aperture of the
corresponding virtual array. We can also see from the figure that
the TS-based MIMO radar with $\zeta = N \lambda/2$ exhibits much
lower stochastic CRB as compared to the case with $\zeta =
\lambda/2$. The reason for this improvement is the larger
effective aperture of the corresponding virtual array. The
traditional MIMO radar with $M$ transmit antennas has the same
effective aperture as that of the TS-based MIMO radar with $\zeta
= N \lambda/2$ but lower SNR per virtual element. Therefore, the
stochastic CRB for traditional MIMO radar is higher than the CRB
for the TS-based MIMO radar with $\zeta = N \lambda/2$. At the
same time, it is better than the CRB of the TS-based MIMO radar
with $\zeta = \lambda/2$ due to larger effective aperture. The
TAP-based MIMO radar has the same effective aperture as that of
the TS-based MIMO radar with $\zeta=\lambda/2$ but higher SNR per
virtual element due to transmit coherent processing gain. It
yields lower CRB. In fact, the CRB for the TAP-based MIMO radar is
comparable to that of the traditional MIMO radar. Finally, the
transmit beamspace MIMO radar with spheroidal sequences based
transmit weight matrix has the lowest CRB as compared to all other
methods. This can be attributed to the fact that the proposed
transmit beamspace-based MIMO radar combines the benefits of
having high SNR due to energy focusing, high power of individual
waveforms, and large effective aperture of the corresponding
virtual array.

\begin{figure}[ht]
\begin{minipage}[b]{1.0\linewidth}
  \centering
  \centerline{\epsfig{figure=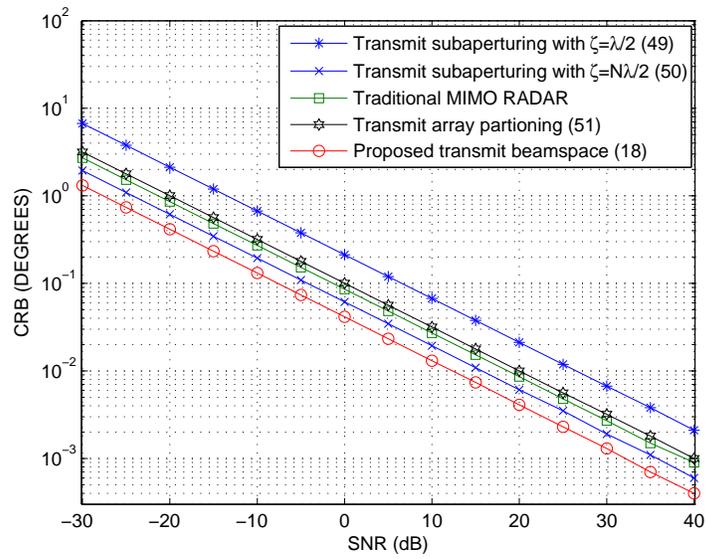,width=10.5cm}}
\end{minipage}
\caption{Deterministic CRB versus SNR.} \label{fig:DetCRB}
\end{figure}

The deterministic CRBs for all methods considered are plotted in
Fig.~\ref{fig:DetCRB}. As can be seen from this figure, the same
observations and conclusion that are drawn from the stochastic CRB
curves also apply to the deterministic CRB.

\subsection{Example~2: MUSIC-based DOA Estimation}
\begin{figure}[ht]
\begin{minipage}[b]{1.0\linewidth}
  \centering
  \centerline{\epsfig{figure=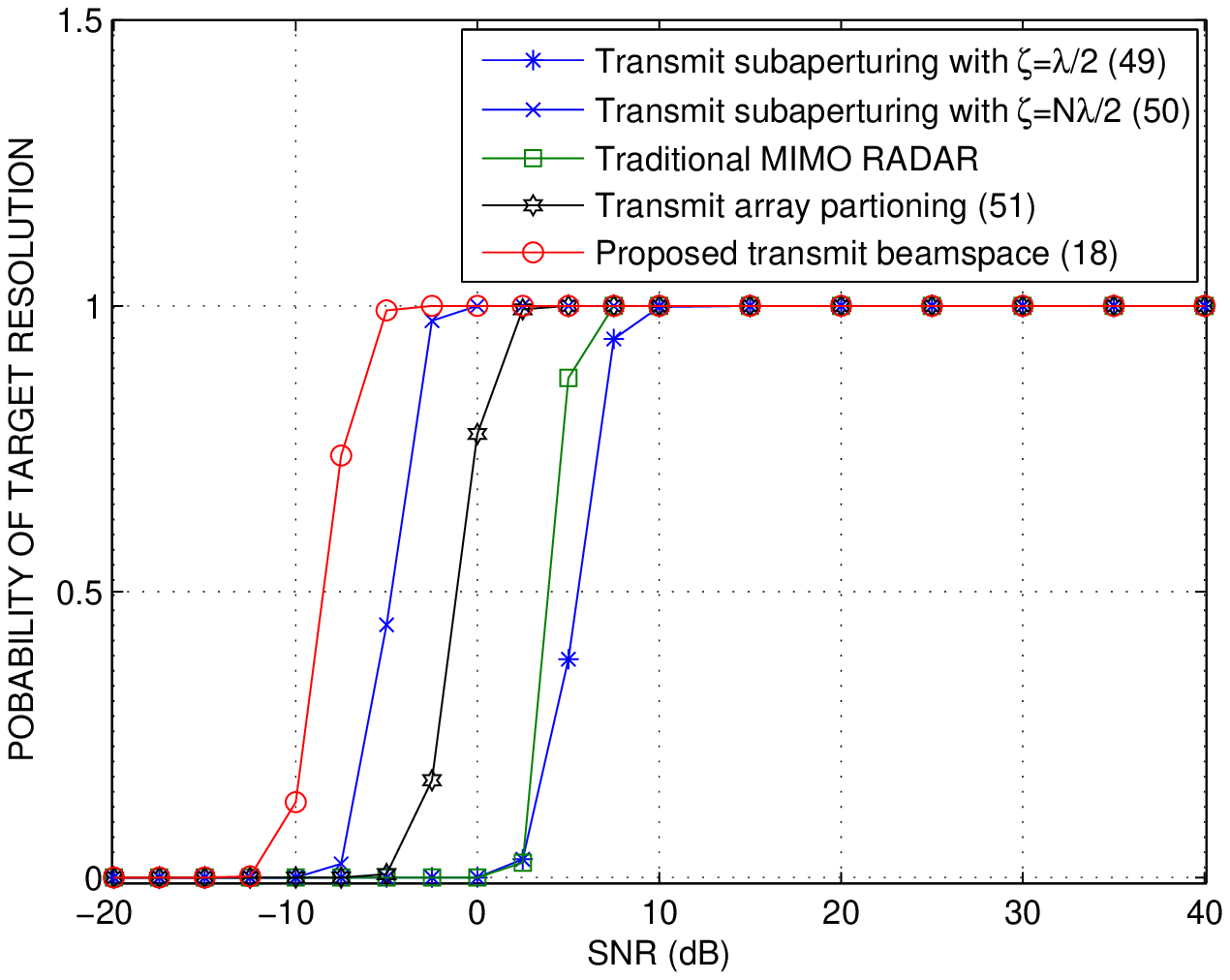,width=10.5cm}}
\end{minipage}
\caption{Probability of target resolution versus SNR for
MUSIC-based DOA estimators.} \label{fig:resolutionMUSIC}
\end{figure}

In this example, the MUSIC algorithm is used to estimate the DOA
for all aforementioned methods. Note that the targets are
considered to be resolved if there are at least two peaks in the
MUSIC spectrum and the following is satisfied \cite{VanTrees}
\begin{equation}\label{eq:resolution}
\left|\hat\theta_l - \theta_l\right|\leq \frac{\Delta\theta}{2},
\quad l=1,2
\end{equation}
where $\Delta\theta = |\theta_2-\theta_1|$.
Fig.~\ref{fig:resolutionMUSIC} shows the probability of source
resolution versus SNR for all methods tested. It can be seen from
this figure that all methods exhibit a $100\%$ correct source
resolution at high SNR values. As the SNR decreases, the
probability of source resolution starts dropping for each method
at a certain point until it eventually becomes zero. The SNR level
at which this transition happens is known as SNR threshold. It can
be seen from Fig.~\ref{fig:resolutionMUSIC} that the TS-based MIMO
radar with $\zeta = \lambda/2$ has the highest SNR threshold while
the traditional MIMO radar and the TAP-based MIMO radar have the
second and third highest SNR thresholds, respectively. The SNR
threshold of the TS-based MIMO radar with $\zeta=N\lambda/2$ is
lower than the aforementioned three methods while the proposed
transmit beamspace-based MIMO radar has the lowest SNR threshold,
i.e., the best probability of source resolution performance.

\begin{figure}[ht]
\begin{minipage}[b]{1.0\linewidth}
  \centering
  \centerline{\epsfig{figure=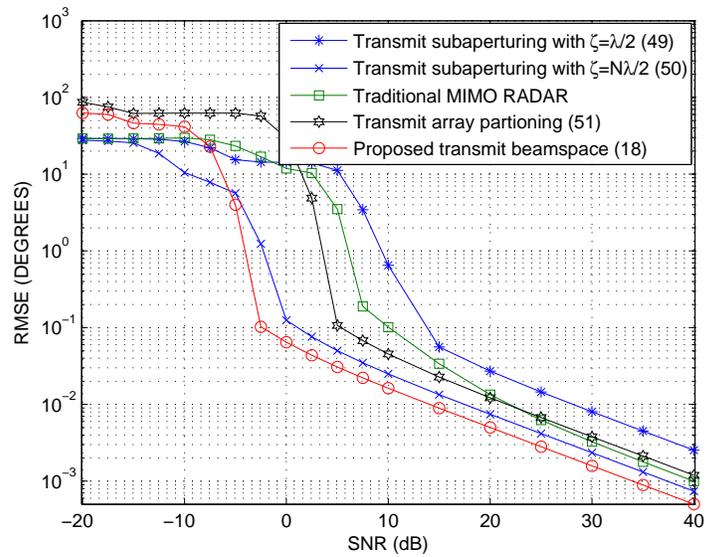,width=10.5cm}}
\end{minipage}
\caption{RMSE versus SNR for MUSIC-based DOA estimators.}
\label{fig:RMSE_MUSIC}
\end{figure}

Fig.~\ref{fig:RMSE_MUSIC} shows the RMSEs for the MUSIC-based DOA
estimators versus SNR for all methods tested. It can be seen from
this figure that the TS-based MIMO radar with $\zeta = \lambda/2$
has the highest/poorest RMSE performance. It can also be seen that
the TAP-based MIMO radar outperforms the traditional MIMO radar at
low SNR region while the opposite occurs at high SNR region. This
means that the influence of having large effective aperture is
prominent at high SNR region, while the benefit of having high SNR
gain per virtual antenna (even if the effective aperture is small)
is feasible at low SNR region. It can also be observed from
Fig.~\ref{fig:RMSE_MUSIC} that the estimation performance of the
TS-based MIMO radar with $\zeta = N\lambda/2$ is better than that
of both the traditional MIMO radar and the TAP-based MIMO radar.
Finally, the proposed transmit beamspace-based MIMO radar
outperforms all aforementioned methods.

It is worth noting that the width of the desired sector is
$10^{\circ}$. Therefore, the parts of the RMSE curves where the
RMSEs exceed $10^{\circ}$ in Fig.~\ref{fig:RMSE_MUSIC} are not
important. Thus, the comparison between different methods within
that region is meaningless.

\subsection{Example~3: ESPRIT-based DOA Estimation}
In this example, all parameters for all methods are the same as in
the previous example except for the $M\times 2$ transmit weight
matrix associated with transmit beamspace-based MIMO radar which
is designed using
\eqref{eq:TransmitOptimization}--\eqref{eq:SidelobeLevels}. The
out-of-sector region is taken as $\bar{\boldsymbol \Theta} =
[-90^{\circ}, \, -15^{\circ}] \cup[15^{\circ}, \, 90^{\circ}]$ and
the parameter that  controls the level of radiation within
$\bar{\boldsymbol \Theta}$ is taken as $\gamma=0.38$. Each column
of the resulting matrix ${\bf C}$ is scaled such that it has unit
norm. The transmit power distribution for this case is similar to
the one shown in Fig.~\ref{fig:pattern2} and, therefore, is not
shown here. ESPRIT-based DOA estimation is performed for all
aforementioned methods. For the traditional MIMO radar-based
method, the $MN\times 1$ virtual array is partitioned into two
overlapped subarrays of size $(M-1)N\times 1$ each, i.e., the
first subarray contains the first $(M-1)N$  elements while the
second subarray contains the last $(M-1)N$ elements. For all other
methods, the $2N\times 1$ virtual array is partitioned into two
non-overlapped subarrays, i.e., the first subarray contains the
first $N$ elements while the second subarray contains the last $N$
elements.

%figure
%
%figure

\begin{figure}[ht]
\begin{minipage}[b]{1.0\linewidth}
  \centering
  \centerline{\epsfig{figure=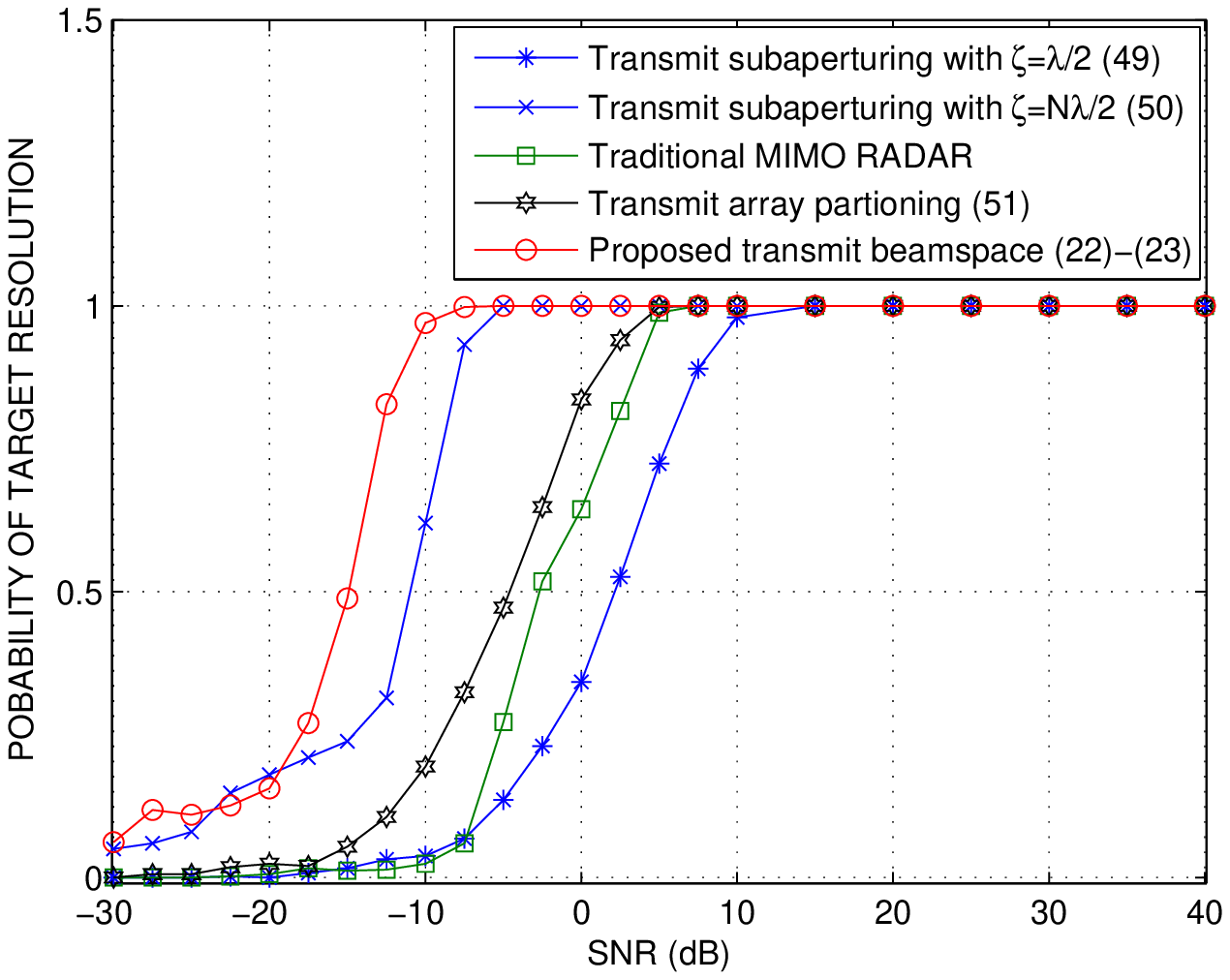,width=10.5cm}}
\end{minipage}
\caption{Probability of target resolution versus SNR for
ESPRIT-based DOA estimators.} \label{fig:resolutionESPRIT}
\end{figure}

\begin{figure}[ht]
\begin{minipage}[b]{1.0\linewidth}
  \centering
  \centerline{\epsfig{figure=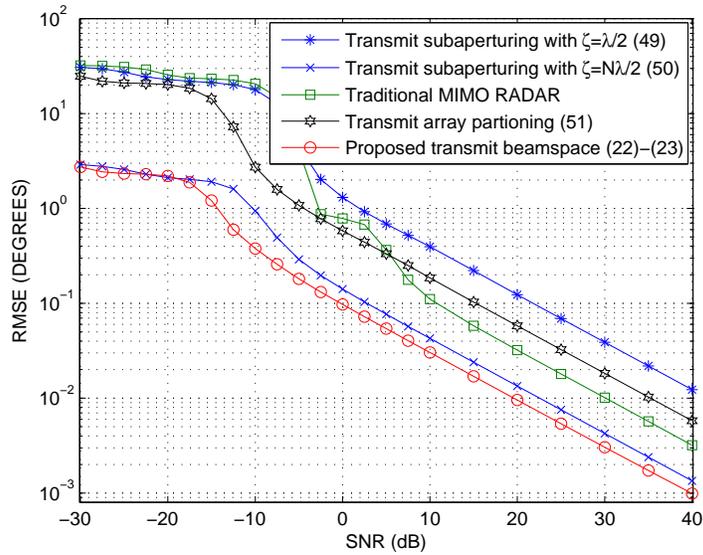,width=10.5cm}}
\end{minipage}
\caption{RMSE versus SNR for ESPRIT-based DOA estimators.}
\label{fig:RMSE_ESPRIT}
\end{figure}

The probability of source resolution and the DOA estimation RMSEs
versus ${\rm SNR}$ are shown for all methods tested in
Figs.~\ref{fig:resolutionESPRIT}~and~\ref{fig:RMSE_ESPRIT},
respectively. It can be seen from Fig.~\ref{fig:resolutionESPRIT}
that the ESPRIT-based DOA estimator for the TS-based MIMO radar
with $\zeta=\lambda/2$ has the highest/poorest SNR threshold while
the ESPRIT-based DOA estimator for the traditional MIMO radar has
the second highest SNR threshold. The ESPRIT-based DOA estimator
for the TAP-based MIMO radar has SNR threshold that is lower than
the previous two estimators. Moreover, the SNR threshold of the
ESPRIT-based DOA estimator for the TS-based MIMO radar with $\zeta
= N \lambda/2$ is lower than the SNR threshold of the ESPRIT-based
DOA estimator for the TAP-based MIMO radar. Finally, the
ESPRIT-based DOA estimator for the proposed transmit
beamspace-based MIMO radar has the lowest SNR threshold, i.e., the
best probability of source resolution performance.

It can be seen from Fig.~\ref{fig:RMSE_ESPRIT} that the
ESPRIT-based DOA estimator for the TS-based MIMO radar with $\zeta
= \lambda/2$ has the highest/poorest RMSE performance. Also this
figure shows that the ESPRIT-based DOA estimator for the TAP-based
MIMO radar outperforms the ESPRIT-based DOA estimator for the
traditional MIMO radar at low SNR region while the opposite occurs
at high SNR region. This confirms again the observation from the
previous example that having large effective aperture is more
important at high SNR region while having high SNR gain per
virtual antenna is more important at low SNR region. It can also
be observed from Fig.~\ref{fig:RMSE_ESPRIT} that the ESPRIT-based
DOA estimator for the TS-based MIMO radar with $\zeta = N
\lambda/2$ outperforms the ESPRIT-based DOA estimator for both the
traditional MIMO radar and the TAP-based MIMO radar. Finally, the
ESPRIT-based DOA estimator for the proposed transmit
beamspace-based MIMO radar outperforms all aforementioned
estimators.

%figure
%figure

\section{Conclusion}
A transmit beamspace energy focusing technique for MIMO radar with
application to direction finding for multiple targets is proposed.
Two methods for focusing the energy of multiple (two or more)
transmitted orthogonal waveforms within a certain spatial sector
are developed. The essence of the first method is to employ
spheroidal sequences for designing transmit beamspace weight
matrix so that the SNR gain at each receive antenna is maximized.
The subspace decomposition-based techniques such as MUSIC can then
be used for direction finding for multiple targets. The second
method uses convex optimization to control the amount of
dissipated energy in the out-of-sector at the transmitter and to
achieve/maintain rotational invariance property at the receiver.
This enables the application of search-free DOA estimation
techniques such as ESPRIT. Performance analysis of the proposed
transmit beamspace-based MIMO radar and comparison to existing
MIMO radar techniques with colocated antennas are given.
Stochastic and deterministic CRB expressions as functions of the
transmit beamspace weight matrix are found. It is shown that the
proposed technique has the lowest CRB as compared to all other
techniques. The computational complexity of the proposed method
can be controlled by selecting the transmit beamspace dimension,
i.e., by selecting the number of transmit beams. Simulation
examples show the superiority of the proposed technique over the
existing techniques.

\newpage

\end{document}